\documentstyle[pre,aps,eqsecnum]{revtex}

\begin{document}

\title{Speckle from phase ordering systems}
\author{Gregory Brown$^{1,2}$ \and Per Arne Rikvold$^{1,2,3}$ 
\and Mark Sutton$^1$ \and Martin Grant$^1$}
\address{
$^1$Physics Department and Centre for the Physics of Materials, 
Rutherford Building, McGill University,
3600 rue University, Montr{\'{e}}al, Qu{\'{e}}bec, Canada H3A 2T8\\
$^2$Center for Materials Research and Technology,
Supercomputer Computations Research Institute,   
and Department of Physics,
Florida State University, Tallahassee, Florida 32306-3016, USA\\
$^3$Department of Fundamental Sciences, Faculty of Integrated Human Studies,
Kyoto University, Kyoto 606, Japan\\
}

\date{\today}
\maketitle

\begin{abstract}
The statistical properties of coherent radiation scattered from
phase-ordering materials are studied in detail using large-scale
computer simulations and analytic arguments.  Specifically, we
consider a two-dimensional model with a nonconserved, scalar order
parameter (Model A), quenched through an order-disorder transition
into the two-phase regime.  For such systems it is well established
that the standard scaling hypothesis applies, consequently the average
scattering intensity at wavevector ${\bf k}$ and time $\tau$ is
proportional to a scaling function which depends only on a rescaled
time, $t \sim |{\bf k}|^2\tau$.  We find that the simulated
intensities are exponentially distributed, and the time-dependent
average is well approximated using a scaling function due to Ohta,
Jasnow, and Kawasaki.  Considering fluctuations around the average
behavior, we find that the covariance of the scattering intensity for
a single wavevector at two different times is proportional to a
scaling function with natural variables $\delta t = |t_1 - t_2|$ and
$\bar{t} = (t_1 + t_2)/2$. In the asymptotic large-$\bar{t}$ limit
this scaling function depends only on $z = \delta t /
\bar{t}\,^{1/2}$.  {}For small values of $z$, the scaling function is
quadratic, corresponding to highly persistent behavior of the
intensity fluctuations.  We empirically establish that the intensity
covariance (for ${\bf k} \neq 0$) equals the square of the spatial
Fourier transform of the two-time, two-point correlation function of
the order parameter.  This connection allows sensitive testing, either
experimental or numerical, of existing theories for two-time
correlations in systems undergoing order-disorder phase
transitions. Comparison between theoretical scaling functions and our
numerical results requires no adjustable parameters.
\end{abstract}

\pacs{
 64.60.My, % Metastable phases
 64.60.Cn, % Order-disorder Transformations
 61.10.Dp, % Theories of X-ray scattering and diffraction
 05.40.+j, % Fluctuation phenom., random process, Brownian motion
}

\section{Introduction} 
\label{sec:Int}

A scattering experiment, using neutrons or X-rays for example, is one
of the most direct measures of the structure of materials. Naively,
this comes about because in the Born approximation, which usually
applies for X-rays and neutrons, the intensity in scattering
measurements is proportional to the Fourier transform of a
density-density correlation function. It is the wavelike properties of
the scattering probe which produces the Fourier transform. For a
deeper understanding of the relationship between scattering intensity
and structure one must realize that this direct correspondence applies
precisely only for coherent waves.  Indeed, for conventional sources,
a given point in the incident wave is only coherent within a
small volume of neighboring points. This coherence volume has
transverse dimensions determined by how parallel the wavefronts are and
a longitudinal length determined by how monochromatic the wave is. In a
standard scattering experiment the different coherence regions of the
incident beam scatter independently.  The intensity measured thus
depends on an incoherent average over different regions of the
scattering volume.  By restricting the scattering volume of the sample
to less than the coherence volume of the beam, one can eliminate this
incoherent average, and thus learn more about the material's
structure.  Of course, the experimental difficulty which arises is
to obtain sufficient diffracted intensity to measure a
signal.  Recently, it has been demonstrated that coherent diffraction
experiments can be performed with X-rays using high-brilliance
synchrotron sources\cite{Sutton:91}.  For coherent diffraction, the
scattering from an inhomogeneous material displays a characteristic
speckled scattering pattern. For instance, the random distribution of
phase-ordering domains shown in Fig.~1 and discussed below, results
in the speckle pattern shown in Fig.~2. As the domains change shape,
the speckle pattern changes, and this time dependence of the speckle
offers a unique method for studying the evolution of inhomogeneous
materials.

Motivated by these advances in experimental techniques, we have
undertaken a theoretical study of intensity fluctuations caused by
scattering from a nonequilibrium system undergoing phase ordering by
domain growth.  When a disordered homogeneous material 
is rapidly brought to a new set of conditions, corresponding to the
coexistence of two equilibrium phases, a spatial pattern of domains of
the two phases develops \cite{Gunton:83}. This change of conditions is often
accomplished by a rapid quench from a high temperature to a low one
below a miscibility gap.  The quench from a homogeneous state with
local fluctuations creates a microstructure of interconnecting,
interlocking domains through the kinetics of a first-order phase
transition. As time goes on the domains grow, so as to minimize the
area of the domain walls that separate the phases. 

When the average domain size $R(\tau)$ at time $\tau$ is large
compared to all other relevant lengths, except for the extent $L$ of
the system itself, the system looks invariant if all lengths are
measured in units of that domain size. In this case the structure of
these many domains is said to ``scale'' with $R(\tau)$.
Experimentally, the growing domain structure is often studied by means
of the scattering intensity \cite{Nagler:88}, whose width is
proportional to the inverse of $R(\tau)$. When the average scattering
intensity is scaled in units of this time-dependent length, one
obtains the scattering function for late times in the form of a
time-independent scaling function.  The time-dependence then enters
only through $R(\tau)$, which can typically be described in terms of
an exponent $n$, such that $R \sim \tau^n$.  This scaling hypothesis
has been found to apply to a large range of systems, and to be
unaffected by many of the microscopic details of specific
materials. That is, the scaling function and the growth exponent $n$
are two features which are common to a large number of systems,
collectively called a universality class.  Universality classes for
phase ordering by domain growth are delineated chiefly by the presence
or absence of conservation laws.  Indeed, many aspects of this
nonequilibrium process can be described by relatively simple
theoretical models.  For example, for systems described by a
nonconserved scalar order parameter, often called Model A
\cite{Gunton:83}, the growth exponent is found to be $n=1/2$.  Systems
included in this class are the Ising model with spin-flip dynamics,
binary alloys undergoing an order-disorder transition, and some
magnetic materials with uniaxial anisotropy.  Model B \cite{Gunton:83}
refers to systems in which the scalar order parameter is conserved,
and the only growth mechanism is diffusion.  Systems in this
universality class have $n=1/3$ and include the conserved Ising model,
as well as binary alloys undergoing phase separation.

In the present paper we investigate the time-dependent fluctuations {\it
around\/} this scaling behavior, and we demonstrate how to study these
fluctuations experimentally through analysis of the time-dependent
scattering. The first theoretical study of such behavior is given in
Ref.~\cite{Roland:89}.  The scattering intensity
is related to the Fourier transform of the order parameter $\psi({\bf
r},\tau)$, the scalar field describing the inhomogeneity of a specific
sample of the scattering material, by
\begin{equation}
\label{eq:intensity}
I({\bf k},\tau)=|\hat\psi({\bf k},\tau)|^2 \;,
\end{equation}
where we ignore the proportionality constant for convenience. 
The average of $I({\bf k},\tau)$ over an ensemble of initial
conditions is the structure factor, 
\begin{equation}
\label{eq:sfactor}
S({\bf k},\tau)=\langle I({\bf k},\tau)\rangle \;.
\end{equation}
Here, the ensemble average expresses the distinction between coherent
scattering, given by $I$, and incoherent scattering, given by
$S$.  Fluctuations around the average scattering are the main topic of
interest in this paper.  The structure factor can also be expressed as
the Fourier transform of the correlation function of $\psi({\bf
r},\tau)$,
\begin{equation}
C({\bf r},\tau) = \langle\psi({\bf 0},\tau)\psi({\bf r},\tau)\rangle \;.
\end{equation}
Because of these relationships, the scattering intensity and structure
factor have been important tools for studying the dynamics of
materials far from equilibrium.  If a system is isotropic, then its
average scattering properties depend only on the magnitude of the
scattering wavevector, $|{\bf k}| = k$.

As mentioned above, scaling by the domain size $R \sim \tau^n$ implies
that correlations are time independent when measured in units of $R$.
Specifically, for an isotropic system, $C(r,\tau) = \overline
C(r/R(\tau))$, where $\overline C (\overline r)$ is one form of the
scaling function.  Fourier transformation gives the average scattering
intensity in terms of another form of the scaling function, which
depends only on a {\it scaled\/} time $t \propto \tau k^{1/n}$,
\begin{equation}
k^d  S(k,\tau)  = F_1(t) \;,
\label{eq:scalt}
\end{equation}
where $d$ is the spatial dimension of the scattering material. The
specific form of $F_1(t)$ depends on the dynamic universality
class.

As discussed above, when the scattering involves {\em coherent} radiation, 
$I({\bf k},\tau)$ is directly measured without
self-averaging.  The scattering from an inhomogeneous
material displays a characteristic speckled scattering pattern much
like the one in Fig.~2, which shows the squared norm of the 
Fourier transform of the configuration in Fig.~1. 
The intensity of each speckle in $I({\bf
k},\tau)$ is due to correlations in the inhomogeneous
sample, and $I - \langle I \rangle = I - S$ gives the fluctuations around the
average scattering intensity.

For materials in equilibrium, deviations of $I$ around the
structure factor are induced by thermal fluctuations
in $\psi$. In fact, fluctuations in the speckle patterns from scattered
laser light have been used as the standard basis of photon correlation
experiments at wavelengths ranging from the ultraviolet to the
infrared \cite{Chu,MandelWolf}. With the advent of high-brilliance 
synchrotron photon sources, experiments involving coherent X-rays have
become possible.  For example, Brownian diffusion rates in gold
colloids have been determined from the time required for a
change in an X-ray speckle pattern \cite{Dierker:95,Chu:95}. 
Speckle from coherent X-rays has also been used to study equilibrium
fluctuations in Fe$_3$Al near an order-disorder transition
\cite{Brauer:95}, 
and in micellar block-copolymer systems \cite{Mochrie:97}. 
X-rays can probe materials on much smaller 
length scales than is currently possible with lasers, and their greater
penetration allows the study of optically opaque materials.

In the present paper, we present a theoretical study of fluctuations
in the scattering intensity from a nonequilibrium system undergoing
phase ordering by domain growth.  For such systems the intensity
fluctuates around a time-dependent structure factor
\cite{Roland:89,Emilio:93,Dufresne:97}. The time evolution of the
intensity at a specific wavevector for one particular quench can be
normalized by the average behavior. A typical example of such a
normalized time series, obtained from a simulation at zero
temperature, is presented in Fig.~3.  It is the correlations of such
time series, averaged over many individual speckles and quenches, that
can be used to study the pattern formation process in phase-ordering
materials.

In Fig.~3 we also show a ``Brownian'' function which was 
constructed to have the same single-time probability density and an
exponential two-time covariance with the same characteristic
time as the normalized nonequilibrium scattering intensity.
Qualitative differences are immediately evident in the two time series.  
The Brownian function fluctuates quickly, with large amplitude 
variations, while the 
intensity fluctuations produced by the phase-ordering system vary
slowly, with markedly less variation in amplitude on short time scales
\cite{Dufresne:97}.  
This property of the nonequilibrium intensity fluctuations is called
persistence \cite{Feder:88}, and it indicates a qualitative difference
between the two-time correlation functions for the two processes \cite{Comm}.

For the remainder of this paper, we specialize to Model A as a simple
model for systems undergoing phase ordering following a quench through
a second-order order--disorder phase transition.  Some preliminary
numerical results from simulations less extensive than the ones used
here were presented in Ref.~\cite{Brown:96}.

The outline of the rest of this paper is as follows.
Section~\ref{sec:meth} describes the details of our numerical
approach, which uses a standard time-dependent Ginzburg-Landau
equation with a nonconserved order parameter.  In Sec.~\ref{sec:corr}
the time-time covariance of individual speckle intensities is
discussed, using analysis readily adaptable to experimental data.  The
equality between this covariance and the square of the two-time
structure factor of the nonequilibrium material is argued in
Sec.~\ref{sec:gauss}, and comparisons to specific two-time theories
follow in Sec.~\ref{sec:theory}. There an analytic expression for the
universal scaling form for the intensity covariance at late times is
obtained.  Finally, our results are summarized in Sec.~\ref{sec:sum}.

\section{Method}
\label{sec:meth}

The dynamics of speckle in ${\bf k}$-space were simulated by
generating successive scattering patterns from a real-space simulation
of the dynamics of phase ordering following a quench through an
order-disorder phase transition. The configuration of the real-space
system is described by a nonconserved scalar order-parameter field
$\psi({\bf r},\tau)$, and its dynamics are governed by the following 
time-dependent Ginzburg-Landau equation,
\begin{equation}
\frac{\partial \psi({\bf r},\tau)}{\partial \tau} = -\Gamma
\frac{\delta {\cal F}[\psi({\bf r},\tau)]}{\delta \psi({\bf r},\tau)}
+ \zeta({\bf r},\tau) \;.
\label{eq:TDGL}
\end{equation}
The first term on the right-hand side of this Langevin equation
corresponds to deterministic relaxation, with rate constant $\Gamma$,
towards a minimum value of the free-energy functional ${\cal
F}[\psi({\bf r},\tau)]$.  Thermal noise, which we neglect, is modeled
by the random variable $\zeta$, whose intensity is proportional to
$\Gamma$ and temperature by virtue of a fluctuation-dissipation
theorem \cite{Gunton:83}.  We neglect $\zeta$ because the most
important sources of noise here are the initial conditions, which give
the random domain morphology.  The main effect of thermal noise at
low temperatures far below any critical point, is only to thermally
roughen the domain walls.

We employ the standard Ginzburg-Landau-Wilson free energy \cite{Gunton:83},
\begin{equation}
\label{Potential}
{\cal F}[\psi({\bf r},\tau)]=\int d{\bf r} \Bigg[-\frac{a}{2}\psi^2({\bf r},\tau)
+\frac{u}{4}\psi^4({\bf r},\tau)
+\frac{c}{2}|{\bf\nabla}\psi({\bf r},\tau)|^2 \Bigg] \;.
\end{equation}
For $a>0,$ the local part of
the integrand in Eq.~(\ref{Potential}) represents a bistable potential,
and the parameters of the model define an equilibrium field magnitude,
$|\psi_0|=\sqrt{a/u}$, 
a thermal correlation length,
$\xi_0=\sqrt{2c/a}$, and a characteristic evolution time,
$\tau_0=(a\Gamma)^{-1}$. The model can be rescaled by choosing a new
field $\tilde\psi=\psi/|\psi_0|$, 
a new length $\tilde{\bf r }=\sqrt{2}{\bf r}/\xi_0$, 
and a new time $\tilde \tau=\tau/\tau_0$. Dropping the tilde from the rescaled
quantities for convenience, we get the rescaled dynamical equation,
\begin{equation}
\frac{\partial\psi({\bf r},\tau)}{\partial \tau}
  = \left( 1 + \nabla^2 \right) \psi({\bf r},\tau) 
  - \psi^3({\bf r},\tau) \;,
\label{eq:EOM} 
\end{equation}
which has no adjustable parameters. Since thermal fluctuations are
explicitly omitted in the simulation, the only randomness comes from
the high-temperature state the system is in before the quench. This was
implemented by an initial condition such that $\psi({\bf r},0)$
consists of independent random numbers uniformly distributed between
$\pm 0.1$.

The simulations were conducted on square lattices with periodic
boundary conditions, lattice constant $\Delta r=1\,$, and a system
size of $L_x=L_y=L=1024$. The Laplacian in Eq.~(\ref{eq:EOM})
was implemented using the eight-neighbor discretization
\cite{Oono:87b,Tomita:91}
\begin{equation}
\label{eq:laplacian}
\nabla^2\psi=\frac{1}{2(\Delta r)^2} \Bigg(\sum \psi_{\rm NN} +
\frac{1}{2} \sum \psi_{\rm NNN} - 6\psi \Bigg) \;,
\end{equation}
where $\psi_{\rm NN}$ are the four nearest neighbors of $\psi$ (along
the lattice directions), 
and $\psi_{\rm NNN}$ are its four next-nearest neighbors
(diagonally).  A simple Euler integration scheme with $\Delta \tau=0.05$
was used to collect data up to a maximum rescaled time of $\tau=2000$ for
$100$ separate sets of initial conditions.

Usually, the order parameter takes the values
$\pm|\psi_0|$ everywhere except at the domain
walls, which are negligibly thin compared to the domains themselves.
After our rescaling, domain walls have a soft nonzero width
of approximately $\sqrt{2}$.  To minimize the effect of this nonzero, though
small, width, we use a nonlinear mapping of the rescaled order parameter to
$\pm 1$ before taking the Fourier transform. The
transformed field, $\hat\psi$, is defined by
\begin{equation} 
\hat\psi({\bf k},\tau) = \frac{1}{\sqrt{L^d}} \sum_{\bf r} 
{\rm sign}\,\Big(\psi({\bf r},\tau)\Big) e^{i{\bf k\cdot r}} \;,
\label{eq:FT} 
\end{equation}
where the fact that the lattice spacing is unity in all directions has
been used. The Brillouin zone in two dimensions is defined by the
discrete set of wavevectors, $k_x$,~$k_y = 2 \pi j/L$ with $j \in \{
0, \, \pm1, \, \pm2, \, \dots, \, \pm(L/2 -1), \, L/2 \}$.  The
scattering intensity $I({\bf k},\tau)=|\hat\psi({\bf k},\tau)|^2$ and
its average over initial conditions is the time dependent structure
factor, $S({\bf k},t)$, as described in Sec.~\ref{sec:Int}. To be
consistent with the numerical integration, the magnitude of the
scattering wavevector, $k({\bf k})$, is defined using the operator
relation
\begin{equation}
-k^2({\bf k}) \hat\psi({\bf k},\tau) = 
\frac{1}{\sqrt{L^d}} 
\sum_{\bf r} e^{i{\bf k \cdot r}} \nabla^2 \psi({\bf r},\tau) \;.
\end{equation}
Substituting the discrete version of the Laplacian from
Eq.~(\ref{eq:laplacian}), one obtains for $d=2$
\begin{equation}
\label{eq:ksquared}
k^2({\bf k})=3-\cos{k_x}-\cos{k_y} 
-\frac{1}{2}\cos{(k_x+k_y)}-\frac{1}{2}\cos{(k_x-k_y)} \;.
\end{equation}
This method has been used to calculate the magnitude of the wavevector
for scaling structure factors and data binning. Because of lattice
effects we consider only those wavevectors with $0<k({\bf k})\le
0.75$.

In order to apply the scaling ansatz, one must know the characteristic
length at time $\tau$.  Often $R(\tau)$ is estimated from the
structure factor, however we have found an analytic expression that
works well. Indeed, one advantage of our numerical approach, as
compared to Monte Carlo or cell dynamical simulations, is that we can
test theories without using any free parameters.  In particular, Ohta,
Jasnow, and Kawasaki \cite{Ohta:82} found that the time dependence of the
domain size obeyed $R(\tau)=\sqrt{4\rho_d \tau}$, with
$\rho_d=(d-1)/d$.  Furthermore, they found that the structure factor
scaled, and they gave an explicit form for the scaling function $F_1(t)$.
In comparison, the theory of Kawasaki, Yalabik, and Gunton
\cite{Kawasaki:78} gives the same form of $F_1(t)$, but the factor
$\rho_d$ does not appear in their result for $R(\tau)$.  Our present
simulations agree with the amplitude given by Ohta {\it et al.\/}, and
so we choose a scaled time $t$ given by 
\begin{equation}
\label{eq:t}
t({\bf k},\tau) = [k R(\tau)]^2 = 4\rho_d k^2 \tau \;.
\end{equation}

\section{Two-time Correlation Functions}
\label{sec:corr}

A quantity to which experiments give ready access is the
fluctuation in the speckle intensity as a function of time.
The relationship between an individual speckle at two different times
$\tau_1$ and $\tau_2$ is, on average, described by the 
intensity covariance,
\begin{equation}
\label{eq:cov}
{\rm Cov_k}({\bf k},\tau_1,\tau_2) =
\Big\langle I({\bf k},\tau_1)I({\bf k},\tau_2)\Big\rangle 
-\Big\langle I({\bf k},\tau_1)\Big\rangle\Big\langle I({\bf k},\tau_2)
\Big\rangle 
\;.
\end{equation} 
For random systems, the covariance is maximum in the equal-time limit,
$\tau_1=\tau_2$, and as the two measurement times become widely
separated, the values of the intensity become stochastically
independent and the covariance decays to zero. For this relaxational
system, negative values are not expected.  The scaling ansatz extended to
this situation allows collapse of the covariance at different $({\bf
k},\tau_1,\tau_2)$ by
\begin{equation}
\label{eq:scalcov}
k^{2d}{\rm Cov_k}(k,\tau_1,\tau_2)={\rm Cov}(t_1,t_2)
\end{equation}
where ${\rm Cov}(t_1,t_2)$ is the scaling function for the covariance.

The simulated scattering intensities were analyzed in the following
way. In Eq.~(\ref{eq:scalcov}), the ensemble average over initial
conditions is scaled to the universal form. To make estimates of ${\rm
Cov}(t_1,t_2)$ from the simulations, we changed the order of the
averaging and used the single-simulation averages $M_1(t)$ and
$M_2(t_1,t_2)$.  $M_1(t)$ is the scaled intensity, $k^d I({\bf
k},\tau)$, averaged over all pairs of $({\bf k},\tau)$ that map onto
$t$, and $M_2(t_1,t_2)$ is the scaled product of the intensities at
two different times, $k^{2d} I({\bf k},\tau_1) I({\bf k},\tau_2)$,
averaged over all triples $({\bf k},\tau_1,\tau_2)$ that map onto
$(t_1,t_2)$.  Due to the large amount of data involved in the present
study, samples for $M_1(t_1)$, $M_1(t_2)$ and $M_2(t_1,t_2)$ were
accumulated in a two-dimensional structure of bins organized by a pair
of variables related to $(t_1,t_2)$ as described below. After
accumulation into the bins $(t_1,t_2)$, averaging over independent
runs, i.e., over different initial conditions, was performed to
further improve our statistics.  These averages were then used to find
the scaling function ${\rm Cov}(t_1,t_2)$.  It should be noted that
all the quantities used here are also readily obtained experimentally,
except for two unknown proportionality constants that depend on
details of the experimental system. One occurs in the equation for the
scaled time, corresponding to $4\rho_d$ in Eq.~(\ref{eq:t}). The other
is the proportionality constant between the measured scattering
intensity and the squared-norm of the Fourier transform of the order
parameter, which we have ignored in Eq.~(\ref{eq:sfactor}). Neither of
these should be a barrier to comparing our results to experiments.

The normalized analog of the covariance is the correlation function
\cite{BrianHappy},
\begin{equation}
\label{eq:corr}
{\rm Corr}({\bf k},\tau_1,\tau_2) = 
\frac
{\Big\langle I({\bf k},\tau_1)I({\bf k},\tau_2)\Big\rangle
-\Big\langle I({\bf k},\tau_1)\Big\rangle\Big\langle I({\bf k},\tau_2)\Big\rangle}
{\sqrt{\Big\langle I^2({\bf k},\tau_1)\Big\rangle-\Big\langle I({\bf k},\tau_1)\Big\rangle^2}
 \sqrt{\Big\langle I^2({\bf k},\tau_2)\Big\rangle-\Big\langle I({\bf
k},\tau_2)\Big\rangle^2}}
\;.
\end{equation}
${\rm Corr}({\bf k},\tau,\tau)$ is unity by construction, which
removes the equal-time variations in the covariance as $\tau$ changes.
Using the definitions of $M_1$ and $M_2$ above, the scaled version can be
expressed as
\begin{equation}
\label{eq:MM}
{\rm Corr}(t_1,t_2) = 
\frac{\Big\langle M_2(t_1,t_2)\Big\rangle
- \Big\langle M_1(t_1) \Big\rangle 
  \Big\langle M_1(t_2) \Big\rangle }
{ \sqrt {
\Big\langle M_2(t_1,t_1)-\big\langle M_1(t_1)\big\rangle^2\Big\rangle
\Big\langle M_2(t_2,t_2)-\big\langle M_1(t_2)\big\rangle^2\Big\rangle
}} \;,
\end{equation}
where $\langle\cdots\rangle$ denotes averaging over initial conditions
as before.
The contour plot of ${\rm Corr}$ in the $(t_1,t_2)$-plane, Fig.~4,
shows how the correlations in individual speckle intensities decay. In
this figure, the line ${\rm Corr}(t_1,t_2)=1$ extends along the
diagonal. Moving away from that line, contours at values of $0.7$,
$0.2$, $0.07$, and $0.02$ are shown. The scatter in the data is
apparent for the last contour and becomes dominant for values of the
correlation less than that. The striking feature of this figure is
that the correlations increase in a nontrivial way as the
phase-ordering continues. Thus, normalization by the time-dependent
intensity is not sufficient to convert the speckle intensity to a
stationary time series.

A more natural set of variables for studying this effect is
$\bar{t}=(t_1+t_2)/2$ and $\delta t=|t_2-t_1|.$ A constant value of
$\bar{t}$ corresponds to a line perpendicular to the $t_1=t_2$
diagonal, while $\delta t$ measures the distance (in units of scaled
time) away from the diagonal. The symmetry under exchange of $t_1$ and
$t_2$ is retained. These variables were used for the binning of
simulation results that produced Fig.~4, with data grouped into $250$
equally wide series for $0\le\bar{t}\le1000$; each $\bar{t}$ series
having $100$ equal-width bins with $0\le\delta t\le2\bar{t}$.

The characteristic time difference, $\delta t_c$, required for the
scaled intensity covariance, ${\rm Cov}(t_1,t_2)$, to decay to half
its maximum value can be found as a function of $\bar{t}$. (The
normalized correlation function, Corr$(t_1,t_2)$, does {\em not\/}
decay to $1/2$ for small values of $\bar{t}$\,.) Here, results for
$M_1$ and $M_2$ for $0\le\bar{t}\le2$ (in $10$ series, with $25$ bins
within each series) were collected in addition to those previously
mentioned. The value of $\delta t_c$ for each $\bar{t}$ was determined
by linear interpolation, and the dependence of $\delta t_c$ on
$\bar{t}$ is presented in log-log form in Fig.~5. In this figure, two
asymptotic limits giving different algebraic relationships are
obvious. For small values of $\bar{t}$ the relationship is linear,
with the exponent determined by least squares being $1.00 \pm 0.01$
for $\bar{t}<1$. At large values of $\bar{t},$ a least-squares fit for
$\bar{t}>200$ gives an exponent of $0.49 \pm 0.02$.  The estimates of
error here are from obtaining the exponents from different ranges of
$\bar{t}$ in the simulation data; statistical error is an order of
magnitude smaller than these estimates.  The exponents we obtain are
in good accord with theories discussed in Sec.~\ref{sec:theory} below,
which give them to be 1 and $1/2$, respectively.  However, the
connection to theory requires some further justification, given in the
following section.

\section{Correlations in the Scattering Material}
\label{sec:gauss}

While experiments can measure time correlations readily through the
covariance or the correlation of intensities, Cov or Corr, theories to
date have made use of the two-point, two-time order-parameter
correlation function, $C({\bf r},\tau_1, \tau_2) \equiv \langle
\psi({\bf 0}, \tau_1)\psi({\bf r}, \tau_2) \rangle$.  In this section
we argue for, and empirically demonstrate, equality between the
intensity covariance for ${\bf k}\neq 0$, which involves fourth
moments of the order parameter, and the square of the spatial Fourier
transform of the two-time order-parameter correlation function in our
model.

The relationship is exact when $\hat\psi({\bf k})$ is a joint Gaussian
random number, with its real and imaginary parts independent and
Gaussian, as we will show below.  However, establishing $\hat\psi({\bf
k})$ to be approximately Gaussian in the present case is not trivial.
Gaussian variables are a natural consequence of the central-limit
theorem, which requires a large number of uncorrelated contributions
to the variable (with some restrictions on the properties of the
individual contributions).  For example, in a disordered system in
equilibrium, correlations exist only on the scale of a small length
$\xi$, so a system of edge-length $L$ consists of on the order 
of $(L/\xi)^d$
independent, uncorrelated parts.  Then the central-limit theorem
applies, and $\hat\psi({\bf k})$ is a complex, Gaussian variable.  
For an ordered system in equilibrium, this argument 
applies to the fluctuations around the
ordered state.  The present nonequilibrium situation has some rough
analogies to a disordered equilibrium state.  At a given time, the
average domain size is $R(\tau)$, and so the number of independent
parts at a given time is approximately $(L/R(\tau))^d$, which can be
large.  It therefore seems reasonable to expect the central-limit
theorem to apply, and indeed, we find empirically that
$\hat\psi({\bf k})$ is Gaussian.  However, unlike a disordered system,
the distribution of domains of different sizes is broad in this case
due to the initial long-wavelength instability and, furthermore, it is
clear that domains interact as they grow.  The degree to which this
correlation is important is a nontrivial issue; below, we test it
numerically and find these correlations to be negligible for the
two-point quantities of interest in this work.

If $\hat\psi({\bf k})$ is Gaussian, 
it is straightforward to relate the intensity 
covariance ${\rm Cov_k}$ to the order-parameter correlation
function $C$.  Wick's theorem can be used to
decompose the intensity-intensity average as
\begin{eqnarray}
\Big\langle I({\bf k},\tau_1)I({\bf k},\tau_2)\Big\rangle & = &
\Big\langle \hat\psi({\bf k},\tau_1)\hat\psi^*({\bf k},\tau_1)
            \hat\psi({\bf k},\tau_2)\hat\psi^*({\bf k},\tau_2) \Big\rangle \\
& = & \quad \hphantom{+} \quad 
      \Big\langle\hat\psi({\bf k},\tau_1)\hat\psi^*({\bf k},\tau_1)\Big\rangle 
      \Big\langle\hat\psi({\bf k},\tau_2)\hat\psi^*({\bf k},\tau_2)\Big\rangle \\
\nonumber
&   & \quad + \quad
      \Big\langle\hat\psi({\bf k},\tau_1)\hat\psi^*({\bf k},\tau_2)\Big\rangle
      \Big\langle\hat\psi^*({\bf k},\tau_1)\hat\psi({\bf k},\tau_2)\Big\rangle \\
\nonumber
&   & \quad + \quad
      \Big\langle\hat\psi({\bf k},\tau_1)\hat\psi({\bf k},\tau_2)\Big\rangle
      \Big\langle\hat\psi^*({\bf k},\tau_1)\hat\psi^*({\bf k},\tau_2)\Big\rangle\\
& = &
\label{eq:correspond}
(1+\delta_{{\bf k},0})S^2({\bf k},\tau_1,\tau_2)
+S({\bf k},\tau_1) S({\bf k},\tau_2) \;.
\end{eqnarray}
Here $S({\bf k},\tau_1,\tau_2)$ is the two-time structure factor
corresponding to the two-point, two-time order-parameter correlation
function
\begin{eqnarray}
\label{eq:skt1t2}
S({\bf k},\tau_1,\tau_2) & = &\Big\langle 
\hat\psi({\bf k},\tau_1)\hat\psi^*({\bf k},\tau_2) \Big\rangle \\
& = & \int d{\bf r} e^{i{\bf k}\cdot{\bf r}} 
C({\bf r},\tau_1,\tau_2) \;.
\end{eqnarray}
For ${\bf k} \neq 0$, Eq.~(\ref{eq:correspond}) can be rewritten as
\begin{equation}
\label{eq:equality}
{\rm Cov_k}({\bf k},\tau_1,\tau_2)=S^2({\bf k},\tau_1,\tau_2) \;,
\end{equation}
which equates the speckle intensity covariance with the square of
the two-time structure factor of the system. Finally, the scaling ansatz
defines a universal two-time form
\begin{equation}
F_2(t_1,t_2)=k^dS(k,\tau_1,\tau_2)
\end{equation}
and
\begin{equation}
{\rm Cov}(t_1,t_2)=F^2_2(t_1,t_2) \;.
\end{equation}

Our numerical tests show that this equality holds well in these
simulations for $k \neq 0$. In computer simulations, unlike scattering
experiments, the two-time structure factor can be found directly using
Eq.~(\ref{eq:skt1t2}). Generally, the product $\hat\psi({\bf
k},\tau_1)\hat\psi^*({\bf k},\tau_2)$ is a complex number, but the
mean value of the imaginary part is zero, so the two-time structure
factor is real valued.  In our simulation data, the imaginary part of
$S(k,\tau_1,\tau_2)$ is found to be zero, within our accuracy.
Results for the real part at $\tau_1=25$ and
$\tau_2=50$ are presented on a log-log scale in Fig.~6.  The direct
measurements of the two-time structure factor and the square-root of
the intensity covariance agree quite well, except at very large
values of $k$, where lattice effects are important.

Another consequence of our proposed decoupling, leading to the
relationship between ${\rm Cov_k}$ and $S$, can be tested by simulation.  When
$\tau_1=\tau_2$, Eq.~(\ref{eq:equality}) is simply
\begin{equation}
\label{eq:var}
 \left\langle I^2({\bf k},\tau)\right\rangle
-\left\langle I({\bf k},\tau)\right\rangle^2
=\left\langle I({\bf k},\tau) \right\rangle^2  \;,
\end{equation}
and scaling can be applied to give 
$F_1(t) = F_2(t,t) = \sqrt{{\rm Cov}(t,t)}$.
These are compared in Fig.~7, where the agreement is clearly quite
good.  As an aside, since the variance does not depend on the system
size, Eq.~(\ref{eq:var}) demonstrates that $I({\bf k},\tau)$ is a
spatially non-self-averaging quantity \cite{Milchev:86}. Because of
this, increasing the system size $L$ will not improve the estimate of
$S({\bf k},\tau)$ given by a specific number of speckles. However,
this is compensated by the fact that more independent speckles are
available in a given range of ${\bf k}$ for each trial. Numerical
results obtained by Shinozaki and Oono in a study of spinodal
decomposition in three dimensions using the cell-dynamical method appear 
to be consistent with this result \cite{Shinozaki:93}. We also note that the
normalization of the two-time intensity correlation function,
Eq.~(\ref{eq:corr}), can be simplified using Eq.~(\ref{eq:var}).

Equation~(\ref{eq:var}) is a property of exponentially distributed
variables, and to the extent that $\hat\psi({\bf k},\tau)$ is a Gaussian
random number, $I({\bf k},\tau)$ will be an exponentially
distributed random number. That is,
the probability density for the {\it normalized} intensities,
$s({\bf k},\tau)={I({\bf k},\tau)}/{S({\bf k},\tau)}$, should satisfy
\begin{equation}
P(s) = \exp{(-s)} \;,
\label{eq:exppdf}
\end{equation}
independent of $({\bf k},\tau)$. The probability density $P(s)$ is
normalized and has unit mean and standard deviation.  Since $P(s)$ is
identical for all values of $({\bf k},\tau)$, only one density
function needs to be constructed. The results for all $0.024<k({\bf
k})<0.75$ are presented for two times in log-linear form in
Fig.~8. The histogram for $s$ is constructed with a bin size of $0.1,$
and the normalized intensity is found using the circular average of
$I({\bf k},\tau)$ from the same trial. The lower bound on $k$ is
chosen such that at least $20$ speckles contribute to this circular 
average. This histogram is accumulated over all $100$ trials and then
normalized; for each histogram on the order of $10^6$ samples are
available. The solid line is the expected density, $P(s)=\exp(-s)$. For
the earlier time the probability density is exponential for all
$s$. At the latest simulation time the density is exponential only for
$s \alt 5$, with a higher than expected occurrence of speckles
brighter than this. Still, less than $0.5\%$ of the points lie this
far into the tail, and the deviation is inferred to be a finite-size
effect because it becomes more pronounced as our simulation
continues. This agrees with earlier simulations \cite{Brown:96}, in
which increasing the system size eliminated the deviation. We believe
this effect is an artifact of the periodic boundary conditions which
allow stabilization of slab-like domain patterns \cite{Wilding:94},
producing ``frozen-in'' interference patterns that remain bright while 
the average intensity steadily decreases.  No $k$ dependence is found
for $P(s).$

\section{Two-time theories}
\label{sec:theory}

Two theoretical predictions for the two-point, two-time
order-parameter correlation function exist, which will be described
and then compared to our simulation results. The first is an analytic
theory, due to Yeung and Jasnow \cite{Yeung:90}, which is an extension
of the analysis by Ohta, Jasnow and Kawasaki \cite{Ohta:82,Ohta:84}. 
The second theory, developed by Liu and Mazenko \cite{Liu:91b}, is
numerical.  Both theories produce approximations for the two-point,
two-time order parameter correlation function, $C({\bf r}, \tau_1 ,
\tau_2)$, in the asymptotic limit.  These then give the structure
factor through the general relation for the Fourier transform of a
spherically symmetric function $f(r)$ in $d$ dimensions,
\begin{equation}
\label{eq:spherFT}
\int d {\bf r} e^{i{\bf k}\cdot{\bf r}} f(r) 
  = (2\pi)^\frac{d}{2} k^{-d} 
  \int_0^\infty d u \, u^\frac{d}{2}\,J_{\frac{d}{2}-1}(u)\, f(u/k)
\;,
\end{equation}
where $J_\nu$ is a Bessel function of the first kind of order $\nu$. 

In the scaling regime, where the domain-wall thickness can be ignored,
the Yeung-Jasnow correlation function is \cite{Yeung:90}, 
\begin{equation}
\label{eq:YJcorr}
C_{\rm YJ}(r, \tau_1, \tau_2) = \frac{2}{\pi} \arcsin 
\left[ \left( \frac{2 R(\tau_1) R(\tau_2)}
                   {R(\tau_1)^2 + R(\tau_2)^2} \right)^{\frac{d}{2}} 
\exp \left( \frac{-r^2}{R(\tau_1)^2 + R(\tau_2)^2}\right) \right] 
\;.
\end{equation}
With the previously defined variables, $\delta t = |t_1 - t_2|$ and 
$\bar{t} = (t_1 + t_2)/2$, this yields 
\begin{equation}
\label{eq:YJdef}
F_{2,\rm YJ}(\delta t,\bar{t}\,)
  =\frac{2}{\pi}(2\pi)^\frac{d}{2} 
  \int_0^\infty du \, u^\frac{d}{2}\,J_{\frac{d}{2}-1}(u)\,
  {\rm arcsin}\Bigg[\Bigg(1-\Big(
              \frac{\delta t}{2\bar{t}}\Big)^2\Bigg)^\frac{d}{4}
  \exp{\Big(\frac{-u^2}{2\bar{t}}\Big)}\Bigg] \;,
\end{equation}

Expansion of ${\rm arcsin}(x)$ about $x=0$, followed by
term-wise integration, gives 
\begin{equation}
\label{eq:YJser}
F_{2,\rm YJ}(\delta t , \bar{t}\,)
=\frac{2}{\pi}(2\pi)^\frac{d}{2}\bar{t}\,^\frac{d}{2}
\sum_{j=0}^{\infty} 
\frac{(2j)!\,\left(1-\left(\frac{\delta t}
                              {2\bar{t}}\right)^2\right)^\frac{d(2j+1)}{4}}
                         {2^{2j}(j!)^2(2j+1)^\frac{d+2}{2}}
\exp{\left(- \frac{\bar{t}}{2(2j+1)}\right)} \;.
\end{equation}
This infinite series is convergent for all physical values of $\delta
t$ and $\bar{t}$; however, its terms are in general nonmonotonic in
$j$.  Only as one or both of the scaled times $t_1$ and $t_2$
approaches zero, is the series well approximated by the $j$=0
term. The asymptotic early-time form,
\begin{mathletters} 
\begin{equation}
\label{eq:YJaxis}
F_{2,\rm YJ}(\delta t,\bar{t}\,) \approx 
\frac{2}{\pi}(2\pi)^\frac{d}{2}
\Bigg(1-\Big(\frac{\delta t}{2\bar{t}}\Big)^2\Bigg)^\frac{d}{4}  
\bar{t}\,^\frac{d}{2}
\exp{\Big(-\frac{\bar{t}}{2}\Big)} 
\;,
\end{equation}
is therefore a good approximation to Eq.~(\ref{eq:YJdef}) 
only in the very restricted region, 
\begin{equation}
\label{eq:YJaxis2}
0 
\leq 
1 - \left( \frac{\delta t}{2 \bar{t}} \right)^2 
\ll 
6^\frac{2}{d} \, 3 \exp \left( \frac{- 2 \bar{t}}{3 d} \right) 
 \;,
\end{equation}
\end{mathletters}
near the $t_1$ and $t_2$ axes or the origin. 

A more useful, analytical 
result is obtained in the limit of large $\bar{t}$ and small $\delta t$. 
Then, the largest terms in the series 
occur in a relatively wide range of $j$ near
$\bar{t}/2(d+3)$. For large $\bar{t}$, the exponential factors 
in Eq.~(\ref{eq:YJser}) 
suppress the small-$j$ terms. The series can then be converted to an
integral, and the factorials can be approximated by Stirling's 
formula to give
\begin{equation}
\label{eq:YJint}
F_{2,\rm YJ}(\delta t ,\bar{t}\,)
\approx \pi^\frac{d-3}{2} 4^\frac{d+1}{2} \bar{t}\,^{-\frac{1}{2}}
\int_0^{\infty} dw \, w^\frac{d-1}{2} 
\left(1 - 
\left(\frac{\delta t}{2\bar{t}}\right)^2\right)^{\frac{\bar{t} d}{8w}} 
\exp{(-w)}
\;.
\end{equation}
The identity $\lim_{m\rightarrow\infty} (1-x/m)^m=\exp(-x)$
is used to obtain the explicitly integrable form, 
\begin{equation}
\label{eq:YJint2}
F_{2,\rm YJ}(z,\bar{t}\,)
\approx \pi^\frac{d-3}{2} 4^\frac{d+1}{2} \bar{t}\,^{-\frac{1}{2}}
\int_0^{\infty} dw \, w^\frac{d-1}{2} 
\exp \left[ - \left(\frac{z^2 d}{32 w} + w \right)\right] 
\;, 
\end{equation}
where $z = \delta t / \sqrt{\bar{t}}$. 
We note that the full Yeung-Jasnow result, 
Eq.~(\ref{eq:YJdef}), 
as well as the early-time approximation, 
Eq.~(\ref{eq:YJaxis}), depends on 
$\delta t$ only through the scaling combination $\delta t / \bar{t}$.
However, in the asymptotic late-time 
approximation, Eq.~(\ref{eq:YJint2}), the natural 
scaling combination is $z = \delta t / \sqrt{\bar{t}}$. 
As we shall see below, this analytical result 
is in excellent agreement with our numerical simulations. 
Equation~(\ref{eq:YJint2}) is readily integrated to yield the explicit 
large-$\bar{t}$ asymptotic scaling function, 
\begin{equation}
\label{eq:YJres}
\bar{t}\,^{\frac{1}{2}}
F_{2,\rm YJ}(z,\bar{t}\,) 
\approx
\pi^\frac{d-3}{2} 2^\frac{d+3}{2} 
\Bigg(z\sqrt{\frac{d}{8}}\Bigg)^\frac{d+1}{2} 
K_{\frac{d+1}{2}}\Bigg(z\sqrt{\frac{d}{8}}\Bigg) \;,
\end{equation}
where $K_n$ is a modified Bessel function of the second kind. 
The right-hand side of this equation depends only on $z$. 
In this large-$\bar{t}$ limit, the asymptotic forms of 
this scaling function with respect to $z$ are
\begin{equation}
\label{eq:YJbigtau}
\bar{t}\,^{\frac{1}{2}}
F_{2,\rm YJ}(z,\bar{t}\,) =
\left\{ \begin{array}{ll}
\pi^\frac{d-3}{2}\Gamma\Big(\frac{d+1}{2}\Big)
4^\frac{d+1}{2}
\Big(1-\frac{1}{16}\frac{d}{d-1}z^2\Big) &{\rm for}\;\; z \ll 1 \\
\pi^\frac{d-2}{2}2^\frac{d+2}{2}
\Big(z\sqrt{\frac{d}{8}}\,\Big)^\frac{d}{2}
\exp{\Big(-z\sqrt{\frac{d}{8}}\,\Big)}
& {\rm for}\;\; 1 \ll z \ll 2\sqrt{\bar{t}}
\end{array} \right. \;.
\end{equation}
The first line of this equation is exact for $z=0$, where it gives the 
asymptotic Porod-tail limit of the Ohta-Jasnow-Kawasaki result for 
$t^{1/2} F_1(t)$.
Note that, in the limit $\delta t=2\bar{t}$, the second line gives a nonzero
value, in contrast to the proper result given in 
Eq.~(\ref{eq:YJaxis}). When $z$ is on the order of $\sqrt{\bar{t}}$,
$F_{2,\rm YJ}$ diverges from the small-$z$ approximation given in 
Eq.~(\ref{eq:YJbigtau}). 

A nonrigorous scaling argument \cite{Yeung:96} suggests that the
order-parameter correlation function should depend on wavevector and
time through $k |R(\tau_1) - R(\tau_2)| \propto \delta t /
\bar{t}\,^{(1-n)}$, when $\delta t \ll \bar{t}$. The scaling variable
$z$ obtained above is consistent with this since $n=1/2$ for Model A.
In fact, the $z=\delta t / \bar{t}\,^{(1-n)}$ result is obtained in the
asymptotic large-$\bar{t}$ limit if one repeats the above calculation
for general $n$, considering the Yeung-Jasnow form for the correlation
function, Eq.~(\ref{eq:YJcorr}), 
simply as an integrable approximation valid for small $r$.

The second theory for two-time correlations in Model A is due to Liu
and Mazenko \cite{Liu:91b}. It is an extension of a theory developed
by Mazenko \cite{Mazenko:90} to predict the universal part of the
two-point, one-time scaled order-parameter 
correlation function $\overline C(r/R(\tau))$. The
heart of the Liu-Mazenko theory is the scaling ansatz
\begin{equation}
\label{eq:lmansatz}
C\Big({\bf r},\tau_1,\tau_2\Big)=
\overline C_{\rm LM}\Big(r/R(\tau_2),\tau_2/\tau_1\Big)
\end{equation}
and the partial differential equation
\begin{equation}
\label{eq:lm}
\frac{\partial \overline C_{\rm LM}({\bf x},\tau')}{\partial \tau'} =
\nabla_{x}^2 \overline C_{\rm LM}({\bf x},\tau') 
+ 2 {\bf x} \cdot {\bf \nabla}_x \overline C_{\rm LM}({\bf x},\tau') 
+ \frac{1}{\mu^*} \tan{\Big(\frac{\pi}{2} 
\overline C_{\rm LM}({\bf x},\tau')\Big)} \;.
\end{equation}
In Eq.~(\ref{eq:lm}), $\overline{C}_{\rm LM}$ is considered a function of 
the rescaled variables ${\bf x}={\bf r}/(2\sqrt{\tau})$ and
$4\tau'=\ln{(\tau_2/\tau_1)}$. Note that these definitions are in
terms of the physical quantities, but the connections to the rescaled
variables used in this work are straightforward. We have found
numerical solutions in terms of Liu and Mazenko's units, which we then
converted so that all results shown here are in terms of the rescaling
given in Section~\ref{sec:meth}. The Mazenko theory for one-time
correlations \cite{Mazenko:90} 
serves to provide input into the Liu-Mazenko model
through the scaled order-parameter 
correlation function 
$\overline C_{\rm M}(x) = \overline C_{\rm LM}(x, \tau'$=0) 
and the numerically obtained 
eigenvalue $\mu^*$.  Since the system is assumed isotropic,
Eq.~(\ref{eq:lm}) can be reduced to a single partial differential
equation in terms of a radial distance $x$ and the logarithmic time
ratio $\tau'$.

The scaled order-parameter 
correlation function in the Liu-Mazenko theory is normalized
to $\overline C_{\rm LM}(0,0)=1$, which causes the tangent term in
Eq.~(\ref{eq:lm}) to diverge. Use of the transformation $G=\sin{(\pi
\overline C_{\rm LM}/2)}$ leads to the small-time solution
\begin{equation}
\overline C_{\rm LM}(0,\tau')=\frac{2}{\pi}{\rm arcsin} {\Bigg(\exp{\Bigg(
-\frac{\pi}{2\mu^*(d-1)}\tau'\Bigg)}\Bigg)} \;.
\end{equation}
This solution is only needed to avoid numerical difficulties at
$x=0$ for the first time increment. Aside from this, Eq.~(\ref{eq:lm})
can be solved numerically using a finite differencing scheme that is
implicit with respect to the derivatives, but evaluates the tangent
term explicitly. Using $\Delta x=0.01\mu^*$ and $\Delta \tau'=10^{-5}$
we have reproduced Fig.~1 of Ref.~\cite{Liu:91b}. In addition, the
exponent for the autocorrelation $\overline C_{\rm LM}(0,\tau')$ (which
will be discussed later) is recovered. The Liu-Mazenko results can be
compared to the theory presented here by taking $\bar{t}$ to be a
parameter and noting that $\delta t=2\bar{t}\tanh{(2\tau')}$. The
two-time structure factor predicted by the Liu-Mazenko theory,
found using the Fourier transform in Eq.~(\ref{eq:spherFT}), is
\begin{eqnarray}
\lefteqn{
F_{2,\rm LM}(\delta t(\bar{t},\tau'),\bar{t}\,) =
 (2\pi)^{\frac{d}{2}}
\Bigg(\frac{2\bar{t}}{\rho_d(1+\exp{(-4\tau')})}\Bigg)^{\frac{d+2}{4}}
\times 
}\\
\nonumber
& &
\int_0^\infty dx \,x^{\frac{d}{2}} J_{\frac{d}{2}-1}
{\Bigg(x\sqrt{\frac{2\bar{t}}{\rho_d(1+\exp{(-4\tau')})}} \Bigg)} 
\overline C_{\rm LM}(x,\tau')
 \;, 
\end{eqnarray}
where the scaling function $\overline C_{\rm LM}(x,\tau')$ itself
depends on $d$.

The two-point, two-time order-parameter 
correlation functions predicted by the two
theories can be compared directly with our simulation, without
adjustable parameters.  Both theories describe the data well in some
instances, and poorly in others.  Our results are presented in
Fig.~9 for $\tau_1=100$ and $\tau_2=200$. Data from our simulation are
circularly averaged with bins of width one in the rescaled distance
units, then averaged over 80 trials. The agreement between the
Yeung-Jasnow theory and our simulation is quite good for this choice
of times, while the Liu-Mazenko result is only qualitatively
correct. 

For larger separations in time, the Yeung-Jasnow theory does
not work as well, most noticeably in predicting the autocorrelation function, 
\begin{equation}
\label{eq:auto}
A(\tau_1,\tau_2) \equiv \langle\psi({\bf r},\tau_1)\psi({\bf r},\tau_2) 
\rangle \;,
\end{equation}
which is equivalent to $C({\bf r=0},\tau_1,\tau_2)$.  
Fisher and Huse
\cite{Fisher:88} have argued that for $\tau_2 \gg \tau_1$ the autocorrelation
obeys the power-law
\begin{equation}
A(\tau_1,\tau_2) \sim [{R(\tau_1)}/{R(\tau_2)}]^\lambda 
\end{equation}
for late times.  
(Note that, since $R\propto \tau^n$, the variables used in this paper give 
$\ln \big[R(\tau_1)/R(\tau_2)\big] =  - 2 n \tanh^{-1} \delta t/2\bar t$.)
Fisher and Huse give physical arguments for $d/2 \le \lambda \le d$.
They point out that $\lambda=1$ for the $d=1$ Glauber model and
conjecture that $\lambda=5/4$ for the two-dimensional spin-flip Ising
model. In the limit $\tau_2\gg\tau_1,$ the Yeung-Jasnow theory gives
$\lambda=d/2,$ as is easily seen by setting $r = 0$ in
Eq.~(\ref{eq:YJcorr}).  The Liu-Mazenko theory yields $\lambda\approx
1.2887$ and $1.6726$ for two and three dimensions, respectively
\cite{Liu:91b}.  Cell-dynamical simulations performed by Liu and
Mazenko \cite{Liu:91b} gave $\lambda=1.246 \pm 0.02$ for $d=2.$ A
recent experiment \cite{Mason:93} found $\lambda=1.246 \pm 0.079$ for
a two-dimensional nematic liquid crystal using video techniques. For
our simulations we measured $A(\tau_1,\tau_2)$ for several values of
$\tau_1.$ The results are presented on a log-log scale versus
$\tau_2/\tau_1$ in Fig.~10. [In this measurement we did not employ the
nonlinear transformation $\psi \rightarrow {\rm sign} (\psi)$. This
accounts for the fact that $A(\tau_1,\tau_1) < 1$, but does not
otherwise seem to affect the results.]  In the figure, the data appear
to support a power-law decay at the latest time, and the exponent
found by fitting the $48$ points $\tau_2>1800$ for $\tau_1=20$ is
$\lambda \approx 1.24$, which is in good agreement with the experiment
and the simulations by Liu and Mazenko. However, the local effective
value of $\lambda$, $\lambda_{\rm eff} = 2 d \ln A / d
\ln(\tau_1/\tau_2)$, which is shown vs. $\tau_2 / \tau_1$ in the inset
in Fig.~10, does not show a clear convergence to an asymptotic limit,
especially for larger values of $\tau_1$. Here $\lambda_{\rm eff}$ is
obtained as a three-point finite-difference estimate around $\tau_2$,
but estimates that smooth the data over wider intervals yield
similarly irregular results.  The estimates of $\lambda$ obtained from
our present simulations are thus somewhat uncertain. However, the
Yeung-Jasnow prediction, $\lambda = 1$, is clearly violated.

The predictions of the two theories for the two-time structure factor
are compared to our simulation data in Fig.~6. The Liu-Mazenko theory
is in semi-quantitative agreement with our simulations at small $k,$
but falls off much more rapidly at large $k$. In addition the shoulder
present both in our simulation result and in the Yeung-Jasnow theory
is absent in the Liu-Mazenko result. The Yeung-Jasnow theory agrees
much better with our simulation, although it does not fall off as fast
at large $k$, and its overestimation around the shoulder is seen
consistently throughout our simulations.

The prediction for the characteristic scaled time separation, $\delta
t_c$, defined in Sec.~\ref{sec:corr}, for both theories is compared to
our simulation results in Fig.~5. All three agree that $\delta t_c =
\tilde D_d \bar{t}$ for small $\bar{t}$. Least-squares fits give
$\tilde D_2=1.36\pm0.02$ for our simulation, $\tilde
D_2=1.325\pm0.002$ and $\tilde D_3=1.633\pm0.006$ for Yeung-Jasnow,
and $\tilde D_2=1.00\pm0.02$ for Liu-Mazenko. At large $\bar{t}$, the
$\delta t_c = D_d \bar{t}\,^{1/2}$ behavior of the Yeung-Jasnow
approach results naturally from Eq.~(\ref{eq:YJres}). This equation
can be solved numerically to find $D_2\approx 2.156$ and $D_3\approx
2.187$ in two and three dimensions, respectively.  Our simulation
results give $D_2=2.12\pm0.01.$ The value of $\bar\tau$ separating the
two scaling behaviors corresponds quantitatively to the value for
which the first term no longer dominates the series expansion of the
Yeung-Jasnow result, given by the relation (\ref{eq:YJaxis2}). The
agreement between the Yeung-Jasnow theory and our simulation is quite
remarkable. On the other hand, for $\bar{t}$ not small the Liu-Mazenko
theory crosses over to a region where $\delta t_c$ is independent of
$\bar{t}$ and then into another linear region; neither of these
relationships is seen in our simulations.

The analytic expression for the universal form of the two-time
structure factor deduced from the Yeung-Jasnow theory for large
$\bar{t}$ is given in Eq.~(\ref{eq:YJres}).  It is tested for several
values of $\bar{t}$ in Fig.~11, which shows good collapse of the
simulation data for $\bar{t} {\rm Cov}(\delta t , \bar{t}\,)$ in terms
of the scaling variable $z=\delta t/\bar{t}\,^{1/2}$. The data also
agree quite well with the Yeung-Jasnow scaling function, $\bar{t}
F^2_{2,{\rm YJ}}$ for $z < 5$.  Indeed, the slow quadratic decay of
correlations near $z = 0$ is another signature of persistence in the
phase-ordering system.  In contrast, Brownian fluctuations give
exponential decay from $\delta t = 0$.  The agreement with the
Yeung-Jasnow theory is remarkable since the scaling of the simulation
data uses an analytic expression for the characteristic length, and no
adjustable parameters are employed.

\section{Conclusions}
\label{sec:sum}

Using both numerical and analytic methods, we have investigated
time-time correlations in the scattering intensity for a
two-dimensional system undergoing an order-disorder transition. The
correlations are found to obey scaling in terms of the variables
$\delta t = |t_2-t_1|$ and $\bar{t}=(t_1+t_2)/2.$ In the large
$\bar{t}$ limit, the correlation data collapse onto a universal curve
which is a function only of $\delta t/\sqrt{\bar{t}}.$

We argue for, and establish numerically, an exponential distribution
for the scattering intensity, Eq.~(\ref{eq:exppdf}), and equality
between the scattering intensity covariance and the square of the
two-time structure factor of the order parameter,
Eq.~(\ref{eq:equality}), for ${\bf k}\neq0$. We use this equality to
test theories for the two-time structure factor due to Yeung and
Jasnow, and Liu and Mazenko.  Both theories describe the data well in
some instances, and poorly in other cases.  The Yeung-Jasnow theory is
very similar to our simulation results, so long as $\delta t$ is not
too large. For $\tau_2 \gg \tau_1$, the Liu-Mazenko theory gives a
better estimate for the autocorrelation exponent $\lambda$.  For large
$\bar{t}$, however, the Liu-Mazenko theory does not show the same
scaling as our simulation results, where the Yeung-Jasnow theory
compares quantitatively well.

Our numerical simulations indicate that a definitive experimental
treatment of time-correlations during an order-disorder transition is
possible by intensity-correlation spectrometry of scattering
speckle. Analysis of experimental correlation data should be similar
to the procedures discussed for the simulation data in
Sec.~\ref{sec:corr}. For non-conserved systems, the experimental
scaling function should be well approximated by Eq.~(\ref{eq:YJaxis}),
for small $k$, with one adjustable parameter for each axis. With a
similar adjustable parameter scheme, the scaling function should be
described by Eq.~(\ref{eq:YJres}) for data in the Porod tail. 

Finally, we expect that the equality between the intensity covariance
and the squared two-time structure factor also occurs in other
phase-ordering systems; in particular, we expect that it occurs for
conserved systems.  That would allow the experimental study of, for
example, time correlations in binary alloys undergoing phase
separation by spinodal decomposition, which are representative of
Model B.  Indeed, preliminary numerical work we have done indicates
this.  Experiments on such systems would be of considerable value.

\section*{Acknowledgments} 
We would like to acknowledge useful discussions with K. Kawasaki,
Y. Oono, M. M. Sano, H. Tomita, and particularly B. Morin and
K. R. Elder.  P.~A.~R.\ is grateful for hospitality and support at
McGill and Kyoto Universities.  Research at McGill University was
supported by the Natural Sciences and Engineering Research Council of
Canada and {\it le Fonds pour la Formation de Chercheurs et l'Aide \`a
la Recherche du Qu\'ebec\/}. Research at Florida State University was
supported by the Center for Materials Research and Technology and by
the Supercomputer Computations Research Institute (under U.S.\
Department of Energy Contract No.\ DE-FC05-85ER25000), and by U.S.\
National Science Foundation grants No.\ DMR-9315969 and DMR-9634873.
P.~A.~R.'s stay at Kyoto University was supported by the Japan
Foundation's Center for Global Partnership through U.~S.\ National
Science Foundation Grant No.\ INT-9512679.  Supercomputer time at the
U.~S.\ National Energy Research Supercomputer Center was made
available by the U.~S.\ Department of Energy.

\newpage

~
\begin{figure}[tbp]
\vskip 2.85in
\includegraphics{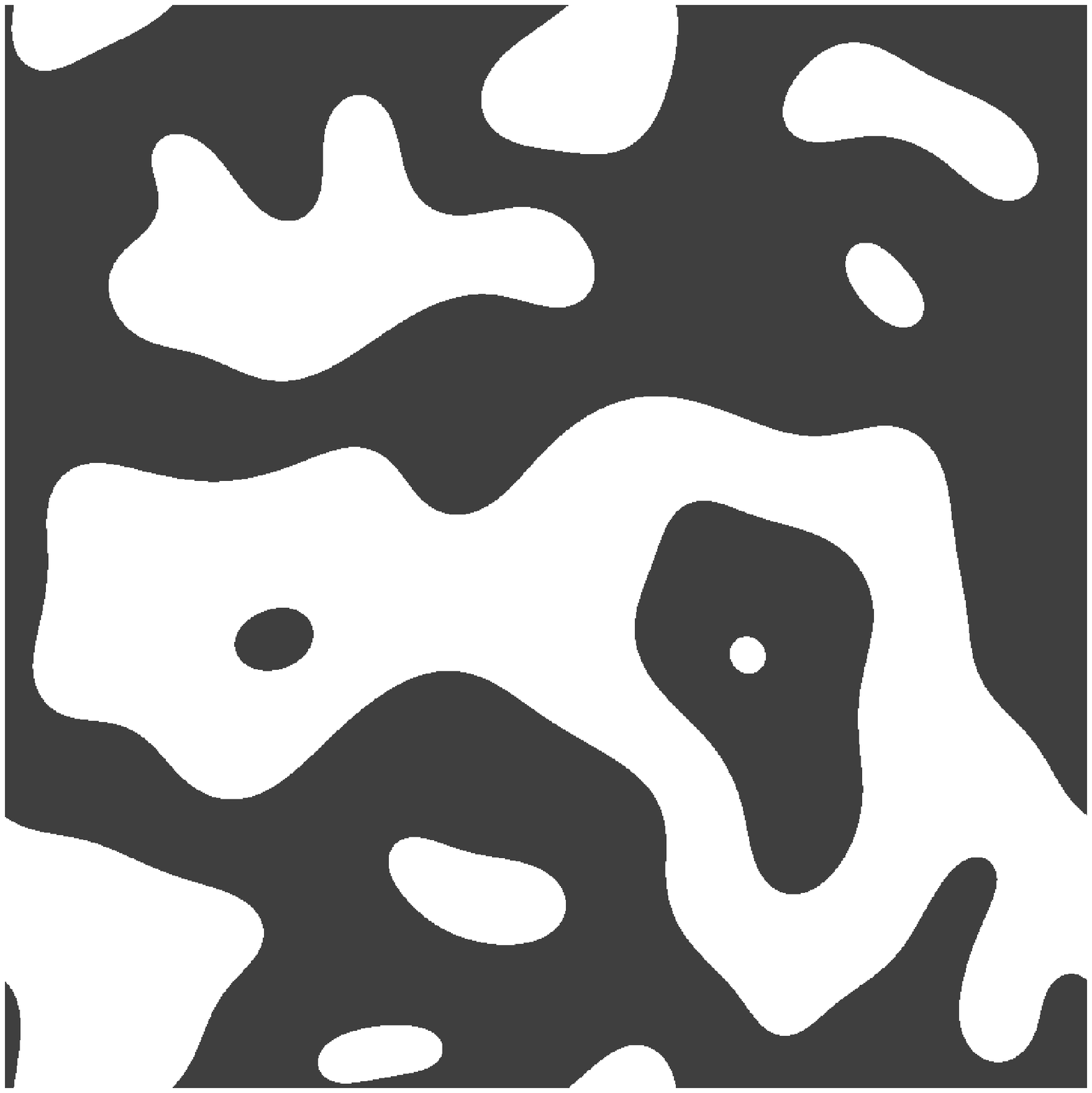}
\bigskip
\caption{A typical configuration of domains, taken from one of the
simulations reported here. Here all systems are $1024 \times 1024$,
and this picture is for the latest simulation time, $\tau=2000$.}
\end{figure}
\vfill

~
\begin{figure}[tbp]
\vskip 2.85in
\includegraphics{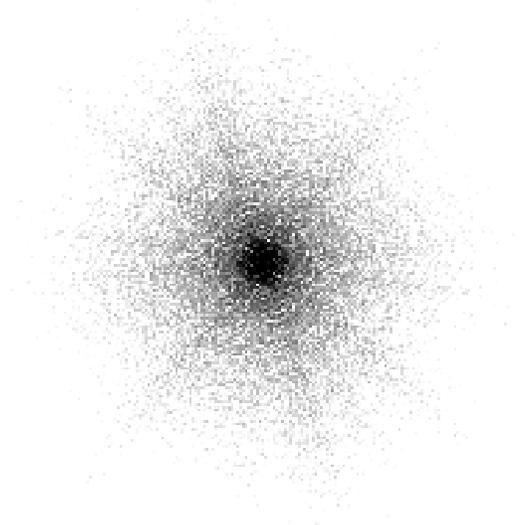}
\bigskip
\caption{Example of the speckled scattering intensity; this
pattern corresponds to the domain structure shown in Fig.~1.
The intensity is shown on a logarithmic scale with darker shades
indicating brighter speckles.  This is a $200 \times 200$ section from
the full $1024 \times 1024$ pattern with the ${\bf k}=0$ origin at the
center of the figure. Speckles do not shift in $\bf k$-space, but
their intensities fluctuate strongly around the $({\bf
k},\tau)$-dependent average value. The rays are present in individual
patterns, but are not correlated between trials.}
\end{figure}
\vfill

~
\begin{figure}[tbp]
\vskip 2.85in
\includegraphics{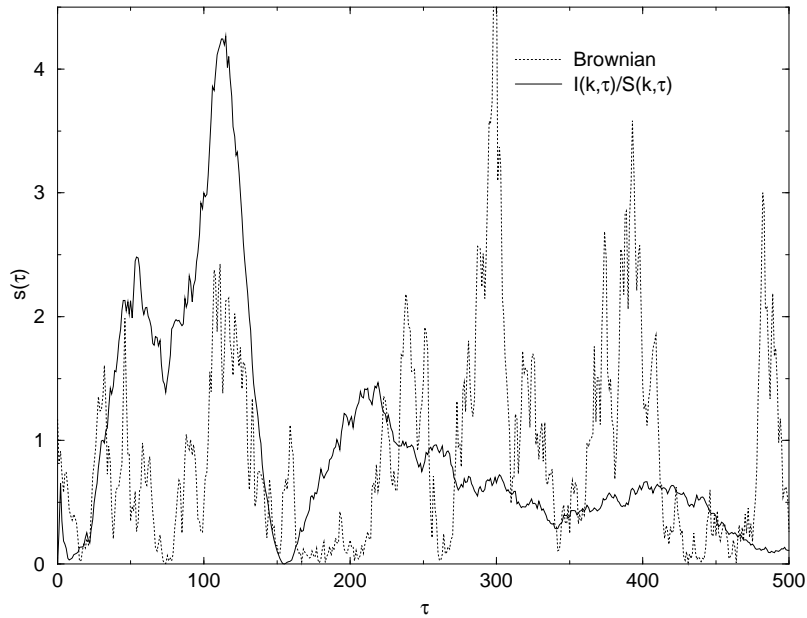}
\bigskip
\caption{Time evolution of the scattering intensity at one
wavevector for one quench to zero temperature. The intensity has been
normalized by the time-dependent structure factor determined over all
100 simulations considered here. The dotted line is a synthetic
``Brownian'' function constructed to have an exponential single-time
probability density and an exponential two-time covariance with a
characteristic time corresponding to that of the simulated
intensity. Compared to the ``Brownian'' function, the persistence of
the scattering intensity is apparent \protect\cite{Feder:88,Comm}.}
\end{figure}
\vfill

~
\begin{figure}[tbp]
\vskip 2.85in
\includegraphics{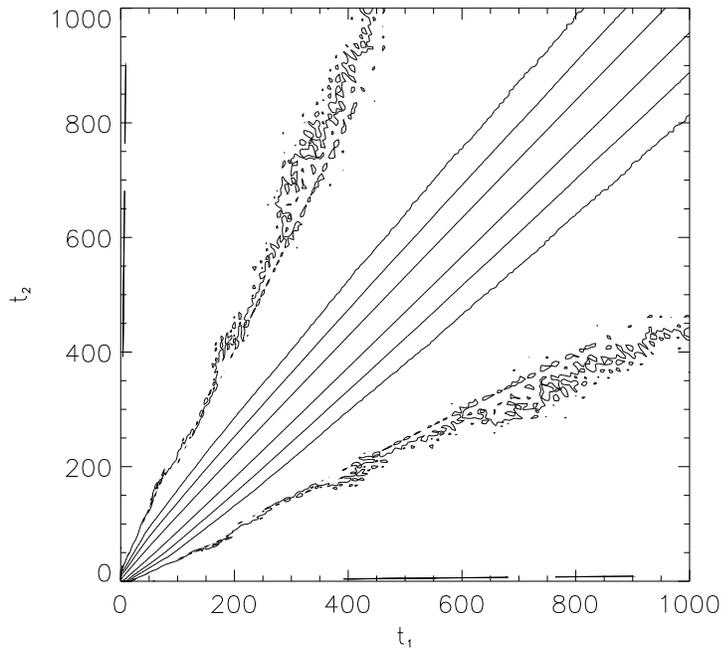}
\bigskip
\caption{Contour plot of the scaled two-time intensity correlation
function, Corr$(t_1,t_2)$. The correlation at $t_1=t_2$ is unity by
construction. The contours moving away from this
diagonal are at $0.7$, $0.2$, $0.07$, and $0.02$. The figure shows
that the speckle intensity stays correlated for larger values of
$\delta t = |t_2-t_1|$ as $\bar{t} = (t_1+t_2)/2$ increases.}
\end{figure}
\vfill

~
\begin{figure}[tbp]
\vskip 2.85in
\includegraphics{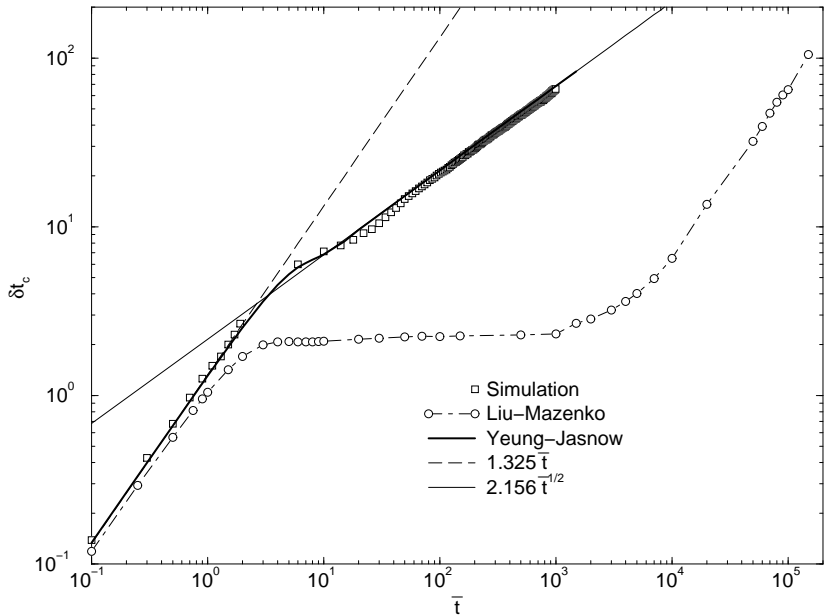}
\bigskip
\caption{The characteristic decay time difference, $\delta t_c,$
as a function of $\bar{t}=(t_1+t_2)/2.$ Power-law behavior is seen at
small and large values of $\bar{t}.$ Least-squares fits to the
simulation data yield a linear relationship at small $\bar{t}$ and
$\delta t_c \sim \bar{t}\,^\alpha$ with $\alpha \approx 1/2$ at large
$\bar{t}$.  The heavy solid line is the analytic prediction of the
Yeung-Jasnow theory. The broken line is its small-$\bar{t}$ limit
$\delta t_c\approx1.325\bar{t}$, and the light solid line is the
large-$\bar{t}$ approximation $\delta t_c \approx
2.156\sqrt{\bar{t}}$. The the $\bar{t}$ associated with the change in
behavior agrees well with Eq.~(\ref{eq:YJaxis2}). The results of the Liu-Mazenko theory are
represented as circles connected by a dot-dashed line; least-squares
fits show them to be nearly linear at both small and large $\bar{t}$.}
\end{figure}
\vfill

~
\begin{figure}[tbp]
\vskip 2.85in
\includegraphics{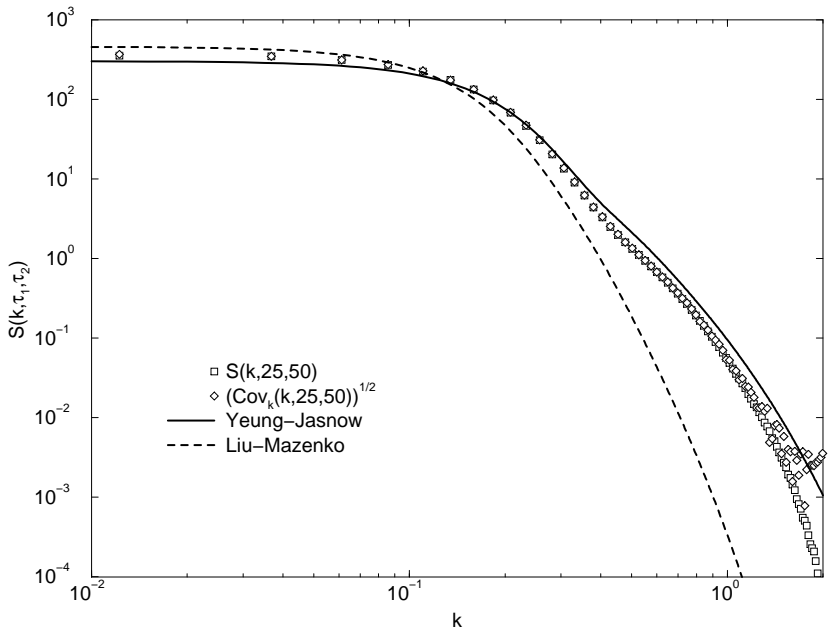}
\bigskip
\caption{The two-time structure factor for $\tau_1=25$ and
$\tau_2=50$. The simulation results are found directly as
$S(k,\tau_1,\tau_2)$ (squares), and indirectly as $\sqrt{{\rm
Cov_k}(k,\tau_1,\tau_2)}$ (diamonds). The Liu-Mazenko theory agrees
semi-quantitatively at small $k.$ The Yeung-Jasnow theory agrees much
better, even reproducing the second shoulder qualitatively.}
\end{figure}
\vfill

~
\begin{figure}[tbp]
\vskip 2.85in
\includegraphics{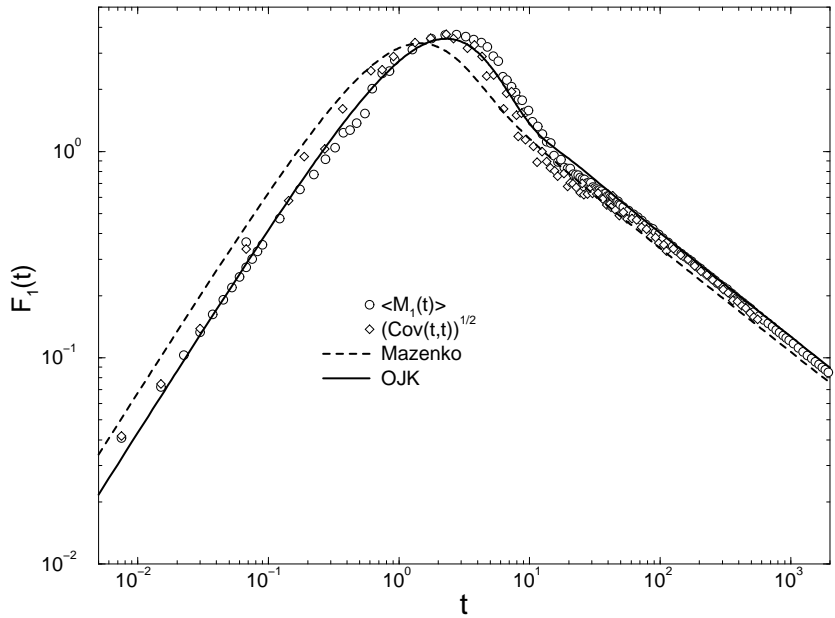}
\bigskip
\caption{Estimates of the scaling function for the structure
factor, $F_1(t)$, obtained from the simulation data using the average
scaled scattering intensity, $\langle M_1(t) \rangle$ (circles), and
the square-root of its variance, $\sqrt{{\rm Cov}(t,t)}$
(diamonds). The agreement between the two measurements is a direct
consequence of the intensity being an exponentially distributed random
variable. The forms obtained from the Yeung-Jasnow (solid line) and
Mazenko (dashed line) theories are also included.}
\end{figure}
\vfill

~
\begin{figure}[tbp]
\vskip 2.85in
\includegraphics{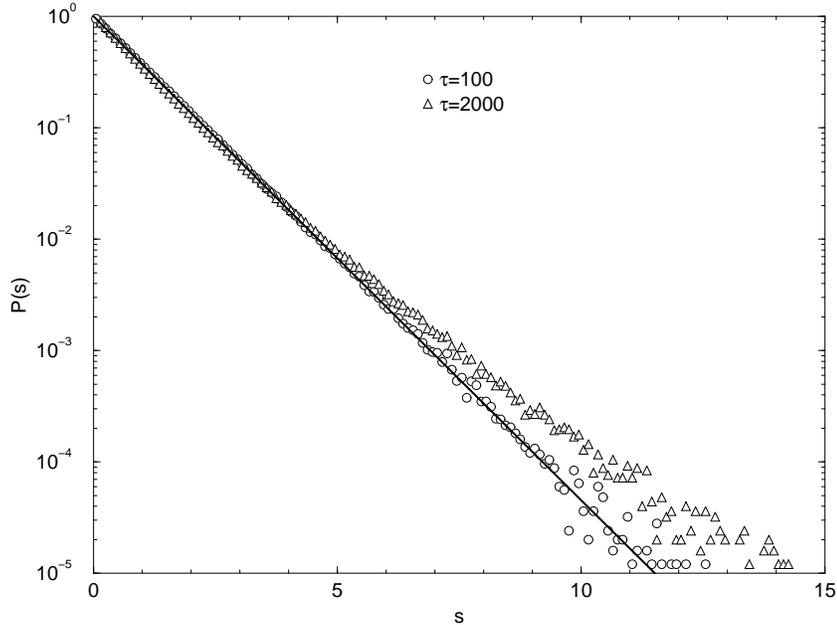}
\bigskip
\caption{The probability density of the normalized speckle
intensity, $s=I({\bf k},\tau)/ S({\bf k},\tau)$, for $\tau=100$ (circles) and
$\tau=2000$ (triangles).  The solid line is the theoretical density of an
exponentially distributed random variable. The deviation at large $s$
of the latest time result is due to finite-size effects.}
\end{figure}
\vfill

~
\begin{figure}[tbp]
\vskip 2.85in
\includegraphics{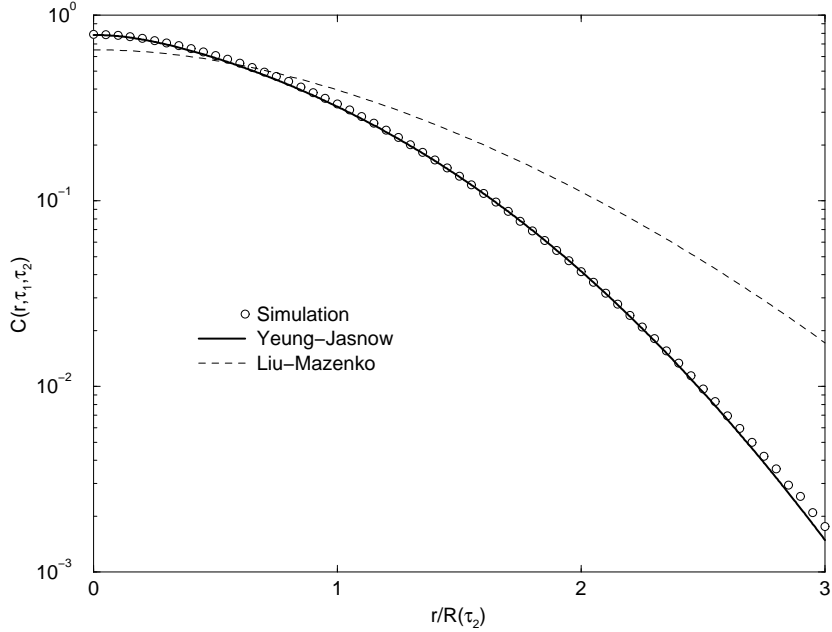}
\bigskip
\caption{The two-time correlation function $C(r,\tau_1,\tau_2)$ with
$\tau_1=100$ and $\tau_2=200$ presented on a semi-log scale. The
characteristic length $R(\tau_2)$ is determined analytically as described in the
text. The agreement between the simulation (circles) and the Yeung-Jasnow theory
(solid line) is quite good for this choice of parameters; the Liu-Mazenko
theory (dashed line) is noticeably different. The agreement between
simulations and Yeung-Jasnow is not as good for larger time
separations.}
\end{figure}
\vfill

~
\begin{figure}[tbp]
\vskip 2.85in
\includegraphics{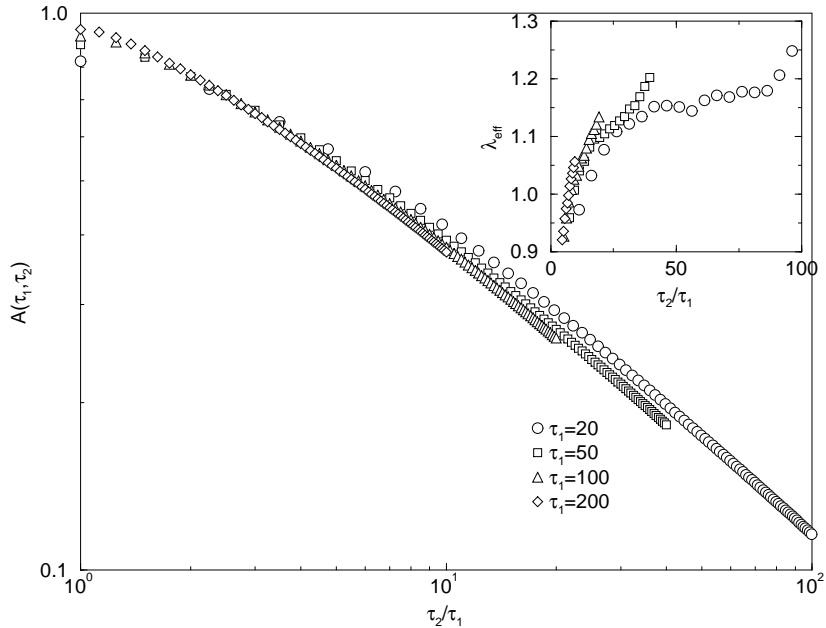}
\bigskip
\caption{The autocorrelation function $A(\tau_1,\tau_2)$. A
least-squares fit for $\tau_2>1800$ for the $\tau_1=20$ data gives the
exponent $\lambda \approx 1.24$.  Fits for later $\tau_1$ give
progressively smaller values, but are still larger than 1. The latest
time value is approached from below, and the asymptotic value may not
be obtained for the simulation times considered here. For two
dimensions, the Liu-Mazenko value of $\lambda$ is found numerically to
be approximately $1.2887$, while a recent experiment gives
$1.246\pm0.079$. The Yeung-Jasnow prediction is unity. The local
effective exponent, $\lambda_{\rm eff}$, estimated from a three-point finite
difference appear in the inset. They characterize
the uncertainty in our estimate of $\lambda$, but show that the
Yeung-Jasnow prediction is violated.}
\end{figure}
\vfill

~
\begin{figure}[tbp]
\vskip 2.85in
\includegraphics{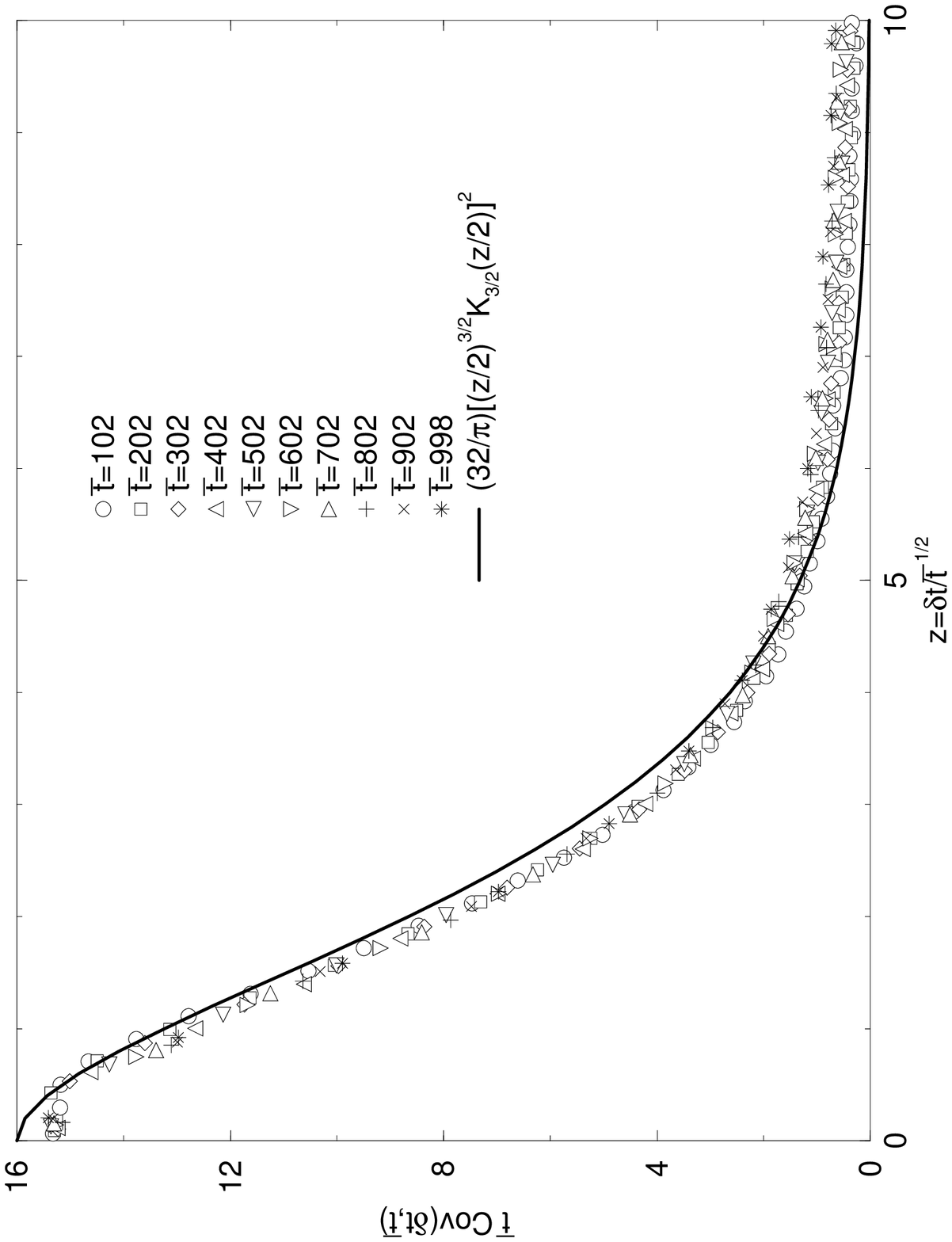}
\bigskip
\caption{Plot of the intensity-covariance scaling function, 
$\bar{t} {\rm Cov}(\delta t , \bar{t}\,)$, 
versus $z = \delta t / \sqrt{\bar{t}}$  
for different values
of $\bar{t}$.  The data collapse onto a single curve
for a large range of $\bar{t}$. The solid curve is the 
corresponding analytic scaling function, 
$\bar{t} F^2_{2,{\rm YJ}}(\delta t , \bar{t}\,)$, 
predicted from the Yeung-Jasnow result in
Eq.~(\protect\ref{eq:YJres}).  It agrees quite well with the
simulation data.}
\end{figure}
\vfill


\begin{thebibliography}{10}
\bibitem{Comm}
A rough, quantitative measure of the different degrees of persistence
can be obtained by application of Range/Scale (R/S) analysis
\protect\cite{Feder:88} to the short sample time series shown in Fig.~3. 
This yields a Hurst exponent of approximately 0.54 for the ``Brownian''
function (near the expected value of 1/2), and a significantly larger
value of approximately 0.74 for the normalized scattering intensity. 
However, the correlation times of these normalized time series are slowly 
increasing. Thus, the processes are not stationary, and the direct 
application of R/S analysis is only to be regarded as an approximate 
way of estimating the degree of persistence.
\bibitem{BrianHappy}
It is natural to normalize the covariance by the standard deviations.
In dynamic light-scattering, the autocorrelation function is normalized
by the average intensity.  For random Gaussian fluctuations,
these are equal.  See Eq.~(\ref{eq:var}).
\end{thebibliography}
\end{document}